\documentclass[twocolumn,aps,amssymb,prl]{revtex4}
\usepackage{graphicx}
\usepackage{epsfig,amsmath}

\begin{document}

\title{Observation of shot-noise-induced asymmetry in
the Coulomb blockaded Josephson junction}

\author{R.~K.~Lindell$^1$, J.~Delahaye$^2$, M.~A.~Sillanp\"a\"a$^1$, T.~T.~Heikkil\"a$^1$,
E.~B.~Sonin$^{1,3}$, and P.~J.~Hakonen$^1$}
\affiliation{ $^1$Low Temperature Laboratory, Helsinki University of
Technology, P.~O.~Box 2200,
FIN-02015 HUT, Finland \\
$^2$LEPES-CNRS, BP 166, 38042 Grenoble Cedex 9, France\\
$^3$The Racah Institute of Physics, The Hebrew University of
Jerusalem, Jerusalem 91904, Israel}

\date{\today} 

\begin{abstract}
We have investigated the influence of shot noise on the IV-curves
of a single mesoscopic Josephson junction.
We observe a linear enhancement of zero-bias conductance of the
Josephson junction with increasing shot noise power. Moreover, the
IV-curves become increasingly asymmetric. Our analysis on the
asymmetry shows that the Coulomb blockade of Cooper pairs is
strongly influenced by the non-Gaussian character of the shot
noise.

\end{abstract}
 \bigskip

\maketitle
Phase coherence of electronic motion is one of the central issues of mesoscopic physics. 
This characteristic is destroyed by environmental noise sources, which are not necessarily
Gaussian. The desire to understand the dephasing mechanisms has given rise 
to a need of mesoscopic noise detectors that can measure the asymmetry
of non-equilibrium noise. Recently, the realization of quantum
noise detectors has been studied theoretically by Aguado and
Kouwenhoven in double quantum dots \cite{Aguado} and by Schoelkopf
et al. in Cooper pair boxes \cite{Schoelkopf}. The possibility of using a
Coulomb blockaded Josephson junction (JJ) as a probe of noise was
shown experimentally \cite{cond-mat}. In addition, the applicability of a SIS mixer for noise detection 
has been demonstrated by Deblock {\it et al} \cite{Deblock}. 

We have investigated the potential of a single mesoscopic JJ as a
detector of shot noise focusing on the case of a strongly resistive
environment, in which Coulomb blockade (CB) of Cooper pair current
\cite{Haviland,Penttila} takes place owing to the delocalization
of the phase variable \cite{SZ}. In this case the current enhances distinctly 
with added non-equilibrium noise \cite{cond-mat}.

The study of higher moments of current fluctuations has recently gained a lot 
of theoretical attention starting with Levitov and Reznikov \cite{Levitov} (see other references in \cite{Reulet,ES}). 
Experimentally, the influence of higher moments has proven to be difficult to study with
standard spectrum analyzer methods. The third moment and asymmetry of shot
noise was, however, measured in this way by Reulet et al \cite{Reulet}.
Tobiska and Nazarov discussed a method for the measurement of higher moments by using 
the asymmetry of escape rates in a current biased, large JJ \cite{Tobiska}.
In our experiments, we generate shot noise by a separately biased
superconductor-insulator-normal metal tunnel junction (SIN TJ). The
quasiparticle current of the TJ is found to strongly reduce the CB of Cooper
pairs: its influence can be resolved down to currents of 0.1 pA. 
We observe a linear enhancement of zero-bias conductance of the Josephson junction with increasing shot noise
power. Moreover, the IV-curves of the JJ become increasingly asymmetric due to the
non-Gaussian nature of shot noise. Our experiment shows, for the first time, 
that the Coulomb blockade of Cooper pairs is strongly influenced by the non-Gaussian nature of 
phase fluctuations generated by shot noise \cite{ES}.

When the supercurrent channel is blocked off owing to a high
impedance environment, the current in a mesoscopic JJ is carried
by incoherent tunneling of Cooper pairs \cite{Aver,IN,IGE}. This is
controlled by the exchange of energy with the
environment. As a result, the junction conductance strongly depends on
temperature: $G_0\propto T^{2\rho -2}$, where the parameter
$\rho=R/R_Q$ is the ratio between the resistance $R$ of the
dissipative ohmic environment and the quantum resistance
$R_Q=h/4e^2$.

The simplest way to account for shot noise is to equate the
available noise power with a noise temperature $T_N$: $2eI_S=4k_B
T_N/R$. Consequently,
the zero-bias conductance $G_0$ under the influence of a
quasiparticle current $I_S$ would correspond to the conductance at an
elevated effective temperature of $T+T_N$. Hence, a comparison of
the current-induced conductance change with the temperature
dependence allows one to determine $T_N$. 

The effective noise temperature analysis has been proven to be
rigorous in the case of small $\rho \ll 1$ \cite{cond-mat}. The theoretical paper of 
Ref.~\onlinecite{ES} shows that such an approach is not valid in the opposite
limit of $\rho \gg 1$. Instead of non-analytic power law
dependence, a linear dependence of $G$ vs $I_S$ is obtained in
the limit $T \rightarrow 0$. 

The IV curve of a Josephson junction at small voltages
can be represented as a power series \cite{ES}
\begin{equation}
   I(V)= I_0 + GV + aV^2 +  b V^3 .
   \label{Shift}
\end{equation}
Here the terms $I_0$ and $aV^2$ are related to the odd moments of shot
noise \cite{ES} and, therefore, they are absent without shot
noise. They characterize the asymmetry of the
$IV$ curves and give rise to the ratchet effect ($I_0$) and a local conductance
maximum. In contrast, the terms $GV$ and $bV^3$ may be present even
without shot noise. In fact, one should write $b = b_0(T) + b_1$, where $b_0$ is a 
temperature dependent term which is present without shot noise and $b_1$ is due to
shot noise.  According to the theory \cite{ES}, the  shot noise contribution 
to the zero bias conductance $G=G_0 +G_S$ is
\begin{eqnarray}
G_S = \frac{\pi ^{5/2}}{32\sqrt{2 \ln \rho}} \left(\frac{E_J}{E_C}\right)^2
\frac{1}{V_C}
\rho ^{3/2} |I_S| \, ,
   \label{G0}
\end{eqnarray}
where $V_C=E_C/\rm{e}=\rm{e}/2 C$. The ratchet current is given by $I_0=\beta_0 I_S$, where

\begin{equation}
   \beta_0 =\frac{\pi ^2}{32}
   \left(\frac{E_J}{E_C}\right)^2 \rho  \, .
   \label{Gain}
\end{equation}
From Eq.~(\ref{Shift}) we also obtain a local conductance extremum (a maximum or 
minimum $-$ depending on the sign of $b$)
\begin{equation}
V_{max} = -\frac{a}{3b} = \rm{sign}(I_S) \frac{2 V_C \ln{\rho}}{\pi^2 \rho},
   \label{vmin}
\end{equation}
where we assumed $b_0 = 0$.
The ratchet effect and the location of the extremum are thus odd functions of $I_S$.

\begin{figure}

    \includegraphics[width=7cm]{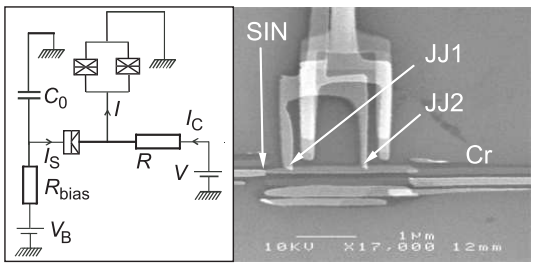}
    \caption{A scanning electron microscope picture of sample 1 and
    a schematic view of the circuit. The chrome resistor is denoted by
    Cr, the superconductor-normal junction by SIN, and the Josephson
    junction in a SQUID-loop configuration by JJ1 and JJ2.}\label{SEM}

    \end{figure}

Our experiments were performed using the circuit layout
displayed in Fig.~\ref{SEM}.  The on-chip structure consists of
three basic elements: 1) an Al-AlO$_x$-Al Josephson junction (JJ), 
2) a superconducting-normal Al-AlO$_x$-Cu tunnel junction (SIN) with, 
(3) a thin film, on-chip, Cr resistor located within a few $\mu$m from 
the Josephson junction. The resistor $R_{bias}$ is located 
outside the cryostat, and it is taken as large as 10 G$\Omega$ 
for good current biasing of the SIN TJ (which has a subgap resistance
$\simeq 1$ G$\Omega$ around zero bias  but $10-100$ M$\Omega$ near the superconducting
gap, which is our typical operating point). In practice, the large capacitance $C_0$
(see Fig.~\ref{SEM}) of the measurement leads 
makes the TJ voltage biased $-$ a necessary condition for shot noise
generation in the junction.
In sample 1, an additional on-chip resistor of 53 k$\Omega$,
was manufactured before the SIN junction. Two samples (see Table I)
with comparable results were investigated. 
The circuits were fabricated using electron beam 
lithography and four-angle evaporation. Linearity of the Cr- resistors was found 
to be better than 5 $\%$, when $B$ and $T$ swept over 0 ... 0.2 T and 0.1 ... 4 K,
respectively.

The Josephson junction was, in fact, made of two $100\times100$ nm$^2$ junctions
in a SQUID geometry in order to enable tuning of its Josephson
energy with an external flux $\Phi$. The Josephson energy $E_J^{max}$
at zero magnetic flux was calculated from the tunneling resistance using the
Ambegaokar-Baratoff relation. The minimum Josephson coupling
energy, $E_J^{min}$, was obtained from the smallest achieved inelastic current
peak in the JJ IV curve. This peak should, according to the $P(E)$-theory \cite{IN},
go down as $\propto E_J^2$, which is also experimentally verified when $E_J < E_C$. 
Intermediate values of $E_J$ can then be determined using the theoretical flux
dependence $E_J=\left[ E_{J_1}^2 + E_{J_2}^2 + 2 E_{J_1} E_{J_2}
\cos(2 \pi \Phi/\Phi_0)\right]^{1/2}$ for the case of asymmetric
Josephson energies $E_{J_1}$ and $E_{J_2}$ for a SQUID.
The Coulomb energy $E_C$ was estimated from the
IV curves in the normal state: the sum of junction capacitances,
$C=C_{SIN}+C_{JJ}$, was obtained from the voltage offset at large
bias voltages using the formula
$V_{\mathrm{offset}}=\frac{e}{2C}$. 

Our samples were cooled with a Leiden Cryogenics dilution refrigerator. The samples 
were mounted inside a tight copper enclosure and the measurement leads were filtered using powder 
rf-filters (Leiden) and 1 m of Thermocoax cable at mixing chamber temperature. 
The JJ current was measured using a DL 1211 current amplifier.

\begin{table}
\begin{tabular}{|c|c|c|c|c|c|}
   \hline  & $R_T^{JJ}(k \Omega )$ & $R_T^{SIN}(k \Omega )$& $R(k \Omega )$
                  & $E_C$  & $E_J^{min}$ / $E_J^{max}$ \\
   \hline 1 & 8.1 & 27.3  & 22.6 & 65 & 22 / 78 \\
   \hline 2 & 24 & 73 & 179 & 80 &  2.4/24  \\

      \hline
  \end{tabular}
  \caption{Sample parameters for two samples numbered by the first column.
  The next two columns give the normal state
  tunneling resistance of the Josephson junction $R_T^{JJ}$ and
  the SIN junction $R_T^{SIN}$. $R$ denotes the
  impedance of the Cr resistors in the immediate vicinity of the Josephson junction.
  The last two columns indicate the Coulomb energy, $E_C$, and the
  minimum $E_J^{min}$ and maximum $E_J^{max}$ values of the Josephson
  energy. The energies are given in $\mu$eV.}

\end{table}

From the measurements of the temperature dependence of $G_0(T)$, we
found that the effective temperature of the sample was between 45$-$80 mK,
somewhat higher than the 25$-$35 mK base temperature of the cryostat.
Nevertheless, as Johnson-Nyquist noise and shot noise are uncorrelated in our case, 
this enhanced thermal noise means only a minor shift in the initial point ($G_0$) of 
the quasiparticle current \cite{quasi} induced conductance changes.

The parameters of sample 2 were most suitable for quantitatively testing the theory of Ref. \onlinecite{ES}.
Therefore, in the following paragraphs, all the results and figures
refer to sample 2.

\begin{figure}

    \includegraphics[width=7.0cm]{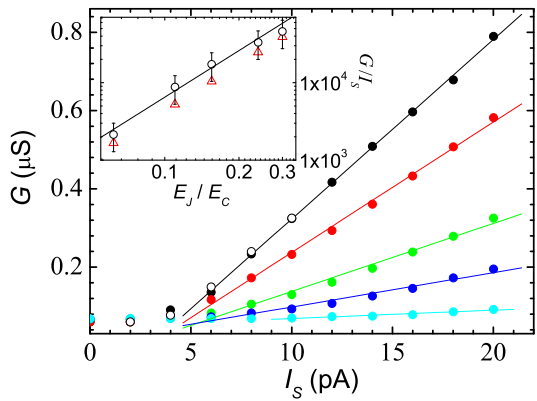}

    \caption{Zero bias conductance for the
    JJ + Cr section of sample 2 as a function of $I_S$ and
    $E_J/E_C$. Beginning from the lowest curve:  $E_J/E_C = $ 0.06, 0.11, 0.16
    0.24 and 0.3. The open circle data points are for $-I_S$. Lines are linear fits to the data. 
    Inset: $G/I_S$ for sample 4 as a
    function of $E_J/E_C$ (open circles). The theoretical values are given by the
    triangles and the fitted line gives an exponent of 2.}\label{G0vsIS}
    \end{figure}

First, we look at the effect of shot-noise power on the conductance
of the JJ, a symmetric effect with respect to sign change of $I_S$.
The zero bias conductance changes substantially  under a small applied current 
through the TJ. This effect is summarized in Fig.~\ref{G0vsIS} for different 
values of the ratio $E_J/E_C$. The linear dependence on $I_S$ is clear (for $I_S >  
5$ pA when $G_S$ becomes comparable to $G_0$), as well as the increase of the
effect with growing $E_J$. Changing the sign of $I_S$ showed indeed that $G_S = G_S(|I_S|)$. 
Since the temperature dependence of
$G_0$ becomes weaker with growing $E_J$, the data of Fig.~\ref{G0vsIS}
also indicate that rising electronic temperature is not
able to explain the change as a function of $I_S$.
In the inset of Fig.~\ref{G0vsIS}, the slopes $G/I_S$ are plotted
together with the predicted $\propto (E_J/E_C)^2$ behavior from Eq.~(\ref{G0}).
The agreement with the theory is within the expected error margin due to  
the large sensitivity of $G_S$ to errors in the experimental determination of $E_J$ and $C$. 

    \begin{figure}

    \includegraphics[width=6.0cm]{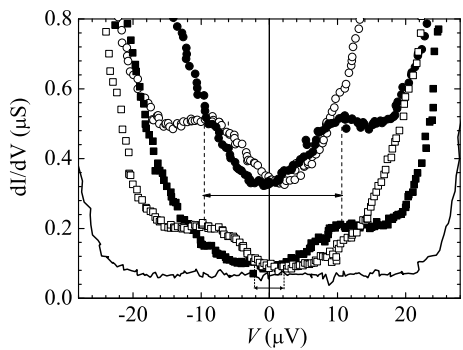}

    \caption{Differential conductance $\frac{dI}{dV}$ for sample 2 for currents:  $I_S = 0$ (solid line), 
    $I_S = 4$ pA ($\blacksquare$), -4 pA ($\square$),
     10 pA ($\bullet$) , -10 pA ($\circ$), 
    The locations of the conductance maxima and minima are indicated by the double arrows.}\label{gv}
    \end{figure}

Next, we consider the asymmetric effects, which are related to the non-Gaussian character of the phase fluctuations
caused by quasiparticles tunneling through the SIN TJ \cite{ES}. In Fig.~\ref{gv}, we show conductance curves
when a small TJ current is applied. One can clearly observe the breaking of the symmetry in the shift of the
conductance minimum as well as through the appearance of a local maximum. The symmetry with respect to sign change of $I_S$ is
also verified by comparing curves with opposite sign of the TJ current. However, the sign of the shift in the minimum conductance
indicates that $b_0 > b_1$, which renders the curvature term $b$ in Eq.~\ref{vmin} positive, and makes the
extremum a minimum. The magnitude of the shift is quite small ($\simeq 2 \mu$eV) as expected from theory. Also, 
the locations of the extrema are independent of $I_S$.

The analytical expansion in Ref.~\onlinecite{ES} found by considering the phase fluctuations over the JJ is only valid near zero 
voltage, and as we move away from that regime additional terms may become significant. Therefore, to explain the appareance and
location of the conductance maximum we have to resort to a qualitative argument, which considers the potential dynamics 
due to quasiparticle tunneling through the SIN junction. When the particle has tunneled, it 
increases the voltage $V_{JJ}$ over the JJ by $\rm{e}/C$ before relaxing after the time $\tau=RC$. 
During this time, the Coulomb blockade of the junction is lifted and current can flow. The effective
voltage over the junction is, however, a time-average of the decaying pulse, and hence
we see the lifting of the blockade at a positive bias voltage \cite{relax}.  
In order to see this effect, a large $RC$ time constant is thus needed to lengthen the non-equilibrium situation 
to the time scale at which inelastic tunneling takes place ($\propto \hbar E_C/E_J^2$ \cite{IN}, which in our
case is 100 ps, while $RC = 180$ ps).

\begin{figure}
        \includegraphics[width=7.0cm]{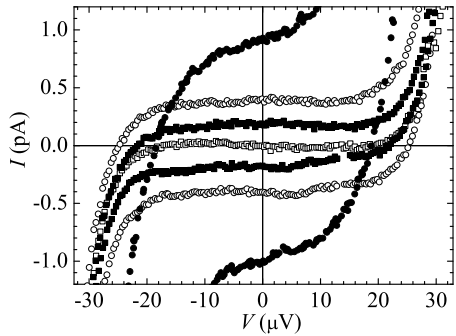}
    \caption{Blockade region IV curves for sample 2, showing the shift of
    IV curve and
    the resulting ratchet effect when the TJ current is $\pm I_S$, where $I_S$ = 0 ($\square$),
    $I_S = 0.2$ pA ($\blacksquare$), $I_S = 0.4 $ pA ($\circ$) and $I_S = 2$ pA 
    ($\bullet$). 
    } \label{ratchet}
    \end{figure}

For sample 2 the ratchet effect could be observed in the IV-characteristics 
at zero voltage. In Fig.~\ref{ratchet}, IV curves of the blockade region are shown 
for small applied TJ currents. The shift of the IV curves and the non-zero current
at zero bias voltage are clearly seen. The fact that the
ratchet effect changes sign for negative $I_S$ shows that the observed shift is not simply due to
current amplifier offset, which is independent of the sign of $I_S$.
The direct measurement was only accurate for small $I_S < 5$ pA, after that the tilting of the conductance curve
started to obscure the ratchet effect at zero voltage. 
The possibility that the non-zero bias gain would be due to a trivial
current division between the chromium resistor $R$ and
the resistance of the JJ, can be ruled out as the Coulomb blockade of the JJ is
quite strong, yielding resistances $>$ 1 M$\Omega$ and a leakage of less than
$0.10 I_S$. 

\begin{figure}

        \includegraphics[width=6.5cm]{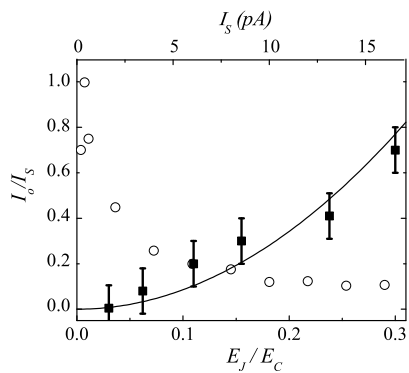}

    \caption{The ratchet current as function of $I_S$ ($\circ$) and $E_J/E_C$ ($\blacksquare$)
    The solid line shows the theoretical prediction with no fitting parameters.}
    \label{ratchet2}
    \end{figure}

In Fig.~\ref{ratchet2} the dependence of $I_0/I_S$ on $I_S$ and $E_J/E_C$ is shown.
The first three points give $I_0 / I_S \simeq 0.7$, while the theoretical gain given by Eq.~(\ref{Gain}) is 0.77. For 
larger values of $I_S$ than 2 pA, $\beta_0$ seems to decline rapidly (roughly as $1/I_S$).
In the same figure, we see that the $E_J/E_C$ dependence of the ratchet effect at small currents ($< 2 $ pA) 
is consistent with the theoretical prediction in Eq.~(\ref{Gain}). 

For sample 1 the asymmetric effects were only observed in the shift of the minimum conductance to finite
currents $I_{JJ}^{\rm min}$, whose sign depends on the sign of $I_S$. From Eq.~(\ref{Shift}) one gets that 
for small currents also this shift, although not strictly at zero voltage, is dominated by the ratchet term. 
We found that this indirectly observed ratchet effect for sample 1 was around 0.8, while the theoretical prediction is 0.5.

According to our analysis, a single Josephson
junction provides a good detector for the total noise power due to 
shot noise. Its sensitivity comes from the
large detector bandwidth, $\sim E_C/h $, foremost owing to the
closeness of the source and the detector. 
According to our experiments, a Coulomb blockaded Josephson
junction can resolve noise from a quasiparticle current as small as
0.1 pA, which clearly exceeds the sensitivity of standard
high-resolution noise experiments \cite{Reznikov}.
%Our current method is highly sensitive to all moments of the underlying
%noise statistics. Therefore, we cannot distinguish individual moments
%in the manner of Ref.~\onlinecite{Reulet}. In order to measure individual moments
%with this method one has to weaken the coupling of shot noise to the 
%detector by reducing $\rho = R/R_Q$.

For the low currents used in this experiment, subsequent voltage pulses
of tunneling quasiparticles do not overlap and, therefore, correlations
between pulses and their statistics are not relevant.
For larger currents ($>$ 100 pA), however, correlations alter the picture and the method
could in principal be used to detect the non-Gaussian electron counting statistics. 

In summary, our measurements of $G$ vs.~$I$ curves for solitary,
resistively confined small Josephson junctions show that the
Cooper pair blockade is a very sensitive probe of not only the total 
noise power but also of the asymmetry of shot noise pulses. In our
measurements, in which the shot noise was induced by quasiparticle current
in a nearby SIN-junction, we find a linear enhancement in the zero bias
conductance with increasing shot noise power as well as asymmetric features
that depend on the sign of the applied current 
The asymmetric effects are consequences of the non-Gaussian phase fluctuations of
shot noise and are in agreement with theoretical predictions.

We acknowledge fruitful discussions with G.~Johansson, L.~Roschier, 
J.~Hassel, H.~Sepp\"a and A.~D.~Zaikin. This work was
supported by the Academy of Finland and by the Large Scale
Installation Program ULTI-3 of the European Union.


\begin{thebibliography}{99}
\bibitem{Aguado} R. Aguado and L. Kouwenhoven, Phys. Rev. Lett. {\bf 84},
1986 (2000).
\bibitem{Schoelkopf} R. Schoelkopf, A. Clerk, S. Girvin, K.
Lehnert, and M. Devoret, cond-mat/0210247.
\bibitem{cond-mat} J. Delahaye, R. Lindell, M. Sillanp\"a\"a, M.
Paalanen, E. Sonin, and P. Hakonen, cond-mat/0209076.
\bibitem{Deblock} R. Deblock, E. Onac, L. Gurevich, and L.P.
Kouwenhoven, Science {\bf 301}, 203 (2003).
\bibitem{Haviland} D. B.  Haviland, L. S. Kuzmin, P. Delsing, and
  T. Claeson, Europhys. Lett. {\bf 16}, 103 (1991).
\bibitem{Penttila} J.S. Penttil\"a, \"U. Parts, P.J. Hakonen, M.A.
Paalanen, and E.B. Sonin, Phys. Rev. Lett. {\bf 82}, 1004 (1999).
\bibitem{SZ} See, {\it e.g.}, G. Sch\"on and A.D. Zaikin, Phys. Rep. {\bf
198}, 237 (1990).
\bibitem{Levitov} L.~S.~Levitov and M.~Reznikov, cond-mat/0111057.
\bibitem{Reulet}B. Reulet, J. Senzier, and D. Prober, Phys. Rev. Lett. {\bf
91}, 196601 (2003).
\bibitem{ES} E. Sonin, cond-mat/0403428.
\bibitem{Tobiska} J. Tobiska and Yu. V. Nazarov, cond-mat/0308310.
\bibitem{Aver} D.V. Averin, Yu.V. Nazarov, and A.A. Odintsov, Physica B
{\bf 165\&166},
945 (1990).
\bibitem{IN} G.L. Ingold and Yu.V. Nazarov, in {\sl Single
Charge Tunneling, Coulomb Blockade Phenomena in Nanostructures},
ed. by H. Grabert and M. Devoret (Plenum, New York, 1992), pp.
21-107.
\bibitem{IGE} G.-L. Ingold, H. Grabert, and U. Eberhardt, Phys.
Rev. B {\bf 50}, 395 (1994).
\bibitem{quasi} The SIN subgap IV curve could readily be fitted with a thermally
activated quasiparticle model. 
\bibitem{relax}
When we take into account the shift of the inelastic tunneling peak from $4E_C$ (-17 $\mu$eV in our case)
due to the finite impedance of the JJ environment and the positive bias voltage at the 
conductance maximum (10 $\mu$eV) we find that the averaging time equals 70 ps, which is consistent
with the time-scale found from $P(E)$-theory (100 ps).
\bibitem{Reznikov} M. Reznikov, R. de Picciotto, T.G. Griffiths,
M. Heiblum, and V. Umansky, Nature {\bf 399}, 238 (1999).
\end{thebibliography}
\end{document}